\documentclass[11pt]{article}

\usepackage{lmodern}             % Use scalable Latin Modern font
\usepackage[margin=1in]{geometry} % Reasonable margins
\usepackage{setspace}
\usepackage{amsmath,amssymb}
\usepackage{graphicx}
\usepackage{subcaption}
\usepackage{float}
\usepackage{titlesec}
\titlespacing{\title}{0pt}{\parskip}{-\parskip}

\usepackage{hyperref}
\hypersetup{
  pdftitle={Modeling Cholera Dynamics with Vaccination as the Control Strategy and Seasonal-forcing Transmission},
  hidelinks,
  pdfcreator={LaTeX via pdfLaTeX}
}

\title{Modeling Cholera Dynamics with Vaccination as the Control Strategy and Seasonal-forcing Transmission}
\author{Eric Herrison Gyamfi}
\date{\vspace{-2.5em}}

\begin{document}
	\maketitle
	
	%\textbf{Abstract:} This study considers \(SVIR-B\) cholera model
	%with impercfect vaccination. We calculate a certain threshold known as
	%instantenuous reproduction number, \(R_v\). If \(R_v <1,\) we obtain a
	%disease-free equilibrium(disease will be elimated from the community)
	%but when \(R_v>1,\) the disease persists in the community. The study
	%perfoms sensitivity analysis of \(R_v\) on two critical
	%values(vaccination rate and wanning rate) in order to determine their
	%relative importance to disease transmission and it is proved an
	%imperfect vaccine is always vital in reducing the disease spread within
	%the community.
	
	\begin{abstract}
		This study presents a seasonally forced \(SVIR\text{-}B\) cholera model that incorporates imperfect vaccination as a control strategy. The model captures the temporal dynamics of susceptible, vaccinated, infected, and recovered individuals, as well as the environmental pathogen concentration. A key focus is the instantaneous reproduction number \(R_v\), which serves as a threshold indicator for outbreak persistence or elimination. When \(R_v < 1\), the disease-free equilibrium is attainable; otherwise, endemic conditions persist. We conduct a sensitivity analysis to evaluate the influence of two critical parameters: the vaccination rate (\(\phi\)) and the waning rate of immunity (\(\theta\)). Results show that increasing the vaccination rate and reducing the waning rate significantly decrease \(R_v\), reinforcing the importance of sustained vaccine efficacy. Seasonal forcing amplifies the complexity of cholera dynamics, revealing the need for timely public health interventions, especially before high-transmission periods. This model demonstrates practical applicability in informing vaccination strategies, especially in resource-limited settings prone to seasonal outbreaks. It offers a flexible framework for public health planning, adaptable to other waterborne diseases. The findings suggest that integrated approaches—combining vaccination, improved sanitation, and targeted education—are essential to reducing cholera transmission and achieving long-term control.
	\end{abstract}

	\section{Introduction}\label{introduction}
	
	According to {[}1{]}, cholera is an acute intestinal infectious disease
	caused by the bacterium \textit{Vibrio cholerae} characterized by
	extreme diarrhea and vomiting. It is deadly water -- borne disease which
	usually results from poor hygienic conditions, sanitation and untreated
	water. The incidence rate is mostly high during raining season of the
	year especially most part of Africa. This is exacerbated by the seasonal fluctuations that heavily influence transmission dynamics, notably during rainy seasons which increase the contamination of water sources . Such seasonal patterns underscore the complexity of cholera transmission, suggesting the importance of incorporating environmental factors into disease modeling and management strategies {[}6, 9{]}. Cholera can either be transmitted through interraction between humans or through interraction between humans and their environment. Individuals who are not treated may die from severe dehydration two or three hours of the infection and this is due to the relatively short incubation period of the disease (usually two to five hours), which will eventually result into an outbreak if it is not controlled and eradicated {[}5{]}.

	While sanitation improvements and hygiene education remain fundamental in combating cholera, these measures alone often fall short in rapidly reducing disease incidence, particularly during outbreaks. Recent epidemiological research emphasizes the need for integrated approaches combining traditional interventions with biomedical solutions such as vaccination {[}1, 8{]}. Incorporating a multidisciplinary approach that leverages insights from medical anthropology alongside public health initiatives could enhance the effectiveness of disease control strategies, particularly by addressing behavioral factors that influence community participation and compliance {[}7{]}.

	% First figure containing c1 and c7
	\begin{figure}[H]
		\centering
		\begin{subfigure}[t]{0.45\linewidth}
			\includegraphics[width=3.5in]{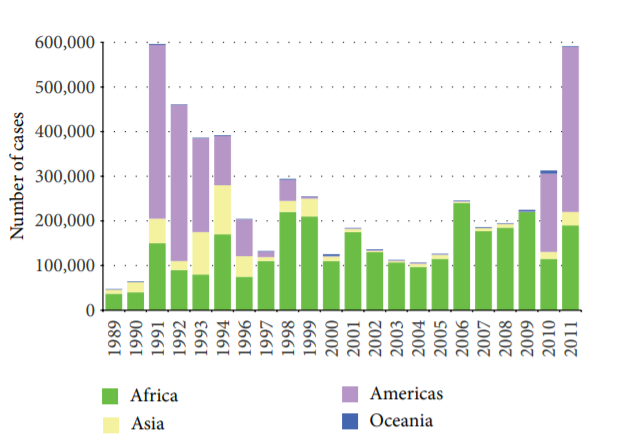}
			\caption{Cholera cases}
			\label{fig:c1}
		\end{subfigure}
		\hfill
		\begin{subfigure}[t]{0.45\linewidth}
			\includegraphics[width=3.5in]{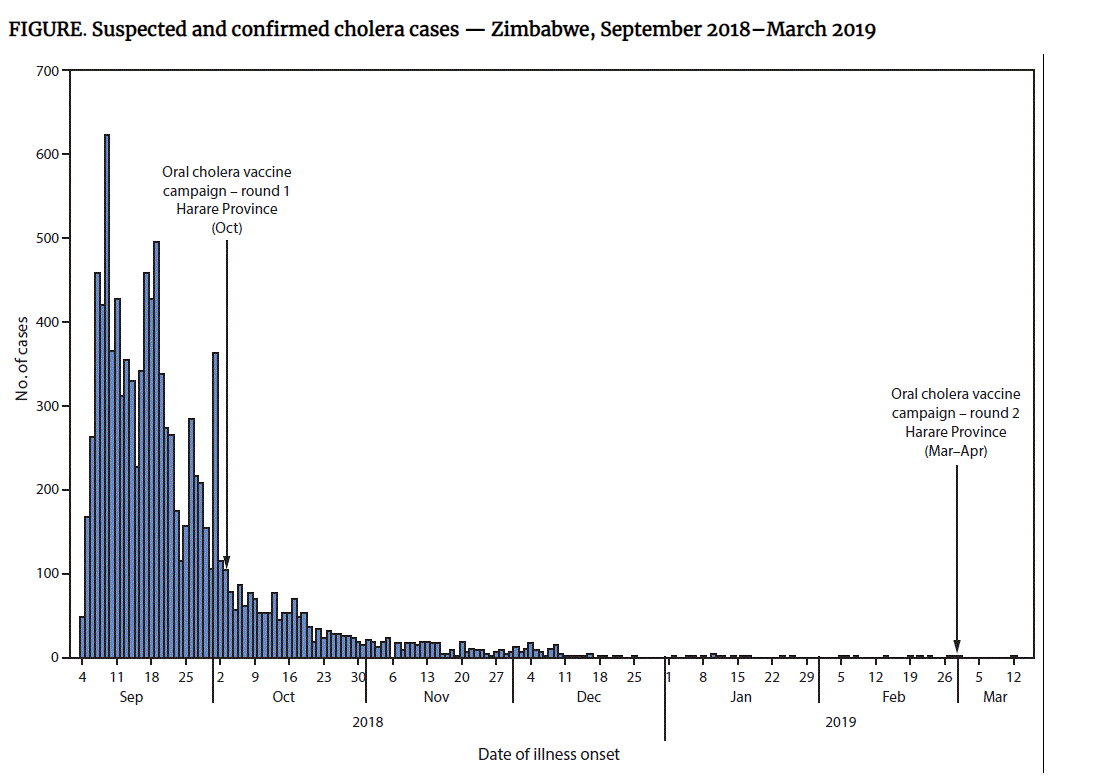}
			\caption{Zimbabwe cases}
			\label{fig:c7}
		\end{subfigure}
		\caption{Cholera cases that have happened over the world}
		\label{fig:c1_c7}
	\end{figure}

	% First figure containing c1 and c7
	\begin{figure}[H]
		\centering
		\begin{subfigure}[t]{0.45\linewidth}
			\includegraphics[width=3.5in]{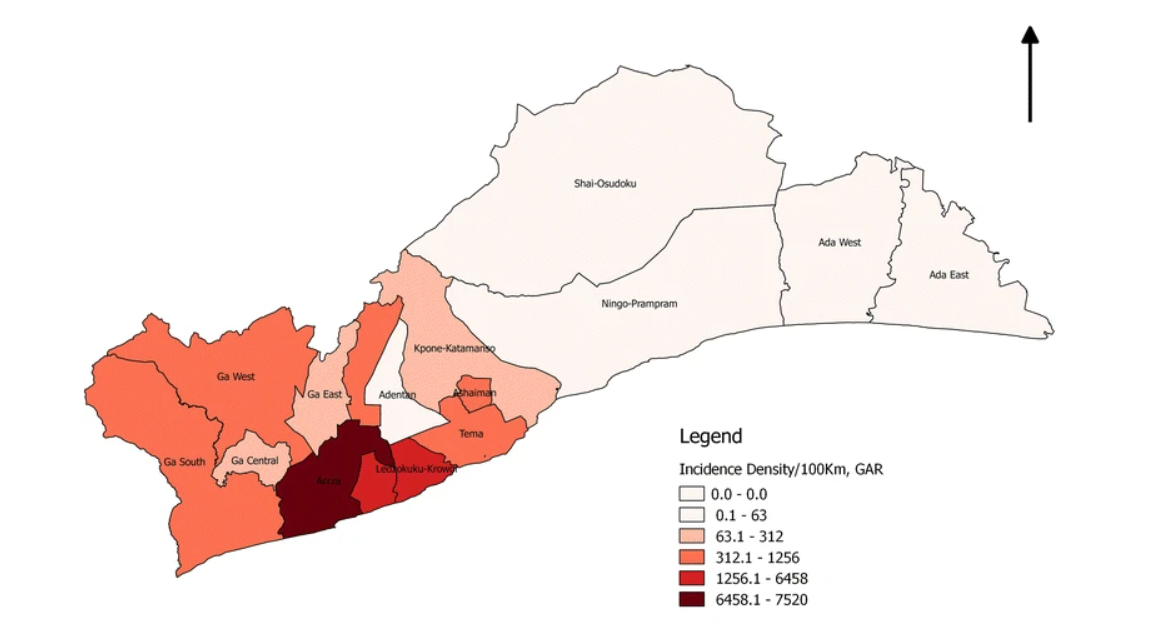}
			\caption{GAR 2014 world cases}
			\label{fig:c2}
		\end{subfigure}
		\hfill
		\begin{subfigure}[t]{0.45\linewidth}
			\includegraphics[width=3.5in]{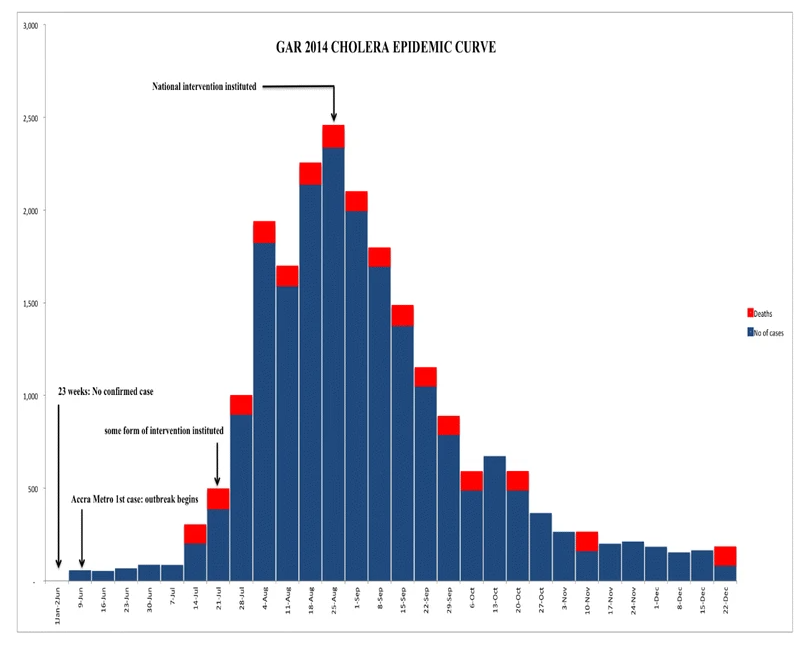}
			\caption{GAR 2014 epidemic curve}
			\label{fig:c4}
		\end{subfigure}
		\caption{Cases that have happened over the world}
		\label{fig:c2_c4}
	\end{figure}

	Globally ( see Figure \ref{fig:c1_c7} and Figure \ref{fig:c2_c4} ), cholera incidence has increased steadily since \(2005\) with cholera outbreaks affecting several continents. This disease continues
	to pose a serious public health problem among developing world populations which have no access to adequate water and sanitation
	resources{[}4{]}. Researchers have developed a variety of vaccines to treat this pathogen. Although vaccination offers a very powerful tool for disease control, generally, vaccines are not \(100\%\) affective and sometimes they only provide limited immunity due to the natural waning
	of immunity in the host. Since Koch found \textit{Vibrio cholera} in 1883, the research for cholera vaccine had been going on for over one
	hundred years. However, these vaccines were parenteral, which have short effective protection and big side effects{[}3{]}. In 1973, the World
	Health Organization canceled the vaccine inoculation which attracted a major concern to oral vaccines. At present, there are three kinds of
	oral vaccines (these includes WC/BS vaccine, WC/rBS vaccine, and CVD103-HgR vaccine) are approved to be safe and effective{[}5{]}. Consequently, ongoing research and mathematical modeling are essential to optimize vaccination strategies by evaluating parameters such as vaccine coverage, efficacy, and immunity duration to achieve maximal public health benefits {[}10, 11{]} .The
	focus of the study is to use epidemilogical and mathematical framework to understand how this imperfect vaccine can help reduce the outbreak
	overtime.
	
	\subsection{Study Contribution}\label{objectives-of-the-study}
	
	The question that is mostly posed in modeling this disease is that,
	``What will be the effect of the vaccine in controlling the disease?''
	and ``What rates in the model have the ability to affect the vaccine?''.
	The study is hypothesized that the vaccination is a good way to control
	cholera, the disease can be controlled if and only if the reproduction
	number is reduced to values less than unity, and waning rate(\(\theta\))
	and vaccinated rate(\(\phi\)) in the model have the ability to affect
	the vaccine. The focus of the study is to:
	
	\begin{enumerate}
		\def\labelenumi{\arabic{enumi}.}
		\item
		understand the dynamics of of response variables over time
		\(S(t), V(t), I(t), R(t)\) and \(B(t)\) over time(exploration
		analysis).
		\item
		perform sensitivity analysis to prove graphically that the critical
		values \(\theta\) and \(\phi\) are important in regulating the
		infection magnitude\((R_v <1)\).
	\end{enumerate}
	
	\section{Methods}
	\subsection{Mathematical Model Formulation}\label{methods}
	
	The study formulates mathematical model using differential equations
	based on the epidemiological compartment modeling. We employ
	\(SVIR-B\)(with demography, \(1=S+V+I+R\)) model for this study, with
	vaccination as a control strategy, which will be incorporated into the
	model. We will numerically simulate data to do some exploration analysis
	and statistical inference. The model assumes seasonal forcing of the
	transmission rate, vaccination leads to death of the pathogen in the
	infected host, human birth and death rate occur at the same rates, and
	this disease occurs in a relatively short period of time and it has low
	mortality.
	
	The model{[}2{]} is given in equation \(1\). The response variables
	\(S(t), V(t), I(t), R(t)\) and \(B(t)\) denote fraction of susceptible,
	infected, vaccinated, recovered individuals and pathogen size at time
	\(t\) respectively. All parameters are assumed nonnegative. The
	parameter \(\mu _1\) represents the natural human birth and death
	rate(same rate for the demography), \(\alpha\) denotes the rate of
	recovery from the disease, \(\eta\) denotes the rate of human
	contribution to the growth of the pathogen, \(\mu _2\) represents the
	death rate of the pathogen in the environment and \(d\) is the
	disease-induced death rate. The coefficients \(\beta _1\) and
	\(\beta _2\) represent the contact rates for the human-environment and
	human-human interactions respectively. The constants \(\alpha _1\) and
	\(\alpha _2\) adjust the appropriate form of the incidence which
	determines the rate of new infection from human and environment. The
	rate at which the susceptible population is vaccinated is \(\phi\), and
	the rate at which the vaccine wears off is \(\theta.\)
	
	If \(\alpha _2 =0,\) then the corresponding incidence is reduced to the
	standard bilinear form and for \(\alpha _2 >0,\) then when the \(I(t)\)
	is high, the incidence rate will respond more slowly than linearly to
	increase in \(I(t)\). \(\alpha _1\) has similar effect to the model. The
	study computes reproduction number, \(R_0\) and the instantaneous
	reproductive number \(R_v\) of \(SVIR-B\) to numerically simulate
	\(S, V, I,R,\) and \(B\). We also analyze the sensitivity of \(R_v\) on
	the two critical values(\(\phi\) and \(\theta\), see figure 3). Figure
	\(1\) is the flow diagram of the model that describes the progression of
	infection from \(S(t)\) and vaccinated \(V(t)\) individuals through the
	\(I(t)\) and recovered \(R(t)\) compartments for the combined
	human-environment epidemiological model with an environmental component.
	
	\begin{figure}[h!]
		
		{\centering \includegraphics[width=6in]{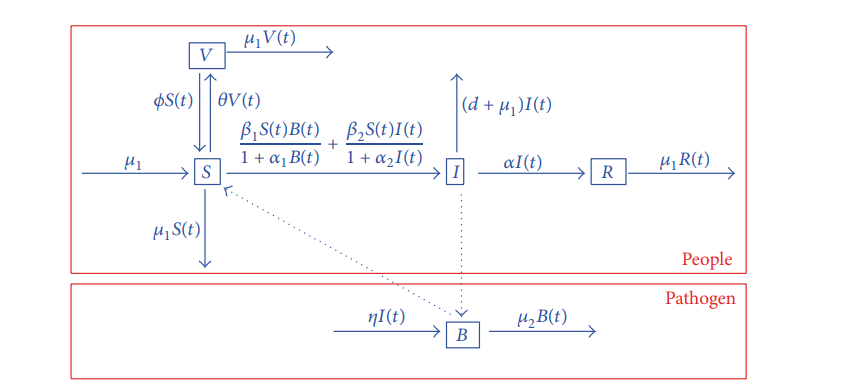} 
			
		}
		
		\caption{The flow diagram of the cholera model.}\label{fig:unnamed-chunk-1}
	\end{figure}
	
	\begin{equation}
	\begin{aligned}
	\frac{dS(t)}{dt} &=  \mu _1 - \frac{\beta _1 S(t)B(t)}{1+\alpha _1B(t)} -\frac{\beta _2 S(t)I(t)}{1+\alpha _2I(t)}-\phi S(t) -\mu _1S(t) +\theta V(t) \\
	\frac{dV(t)}{dt} &=  \phi S(t) - \theta V(t) -\mu _1 V(t)\\
	\frac{dI(t)}{dt} &=  \frac{\beta _1S(t)B(t)}{1+\alpha _1B(t)} + \frac{\beta _2S(t)I(t)}{1+\alpha _2I(t)}-(d +\alpha + \mu _1) I(t) \\
	\frac{dR(t)}{dt} &= \alpha I(t) - \mu _1 R(t)\\
	\frac{dB(t)}{dt} &= \eta I(t) - \mu _2 B(t)
	\end{aligned}
	\end{equation}
	
	From literature, we obtain the seasonal forcing parameters
	\(\beta _1 =\beta _{01}\bigg(1 + a*cos(2\pi t)\bigg)\) and
	\(\beta _2 =\beta _{02}\bigg(1 + b*cos(2\pi t)\bigg)\) where \(a\) and
	\(b\) denote the rate of the seaonal forcing, \(\beta _{01}\) and
	\(\beta _{02}\) represent the initial values of the transmission rate
	when there is no seasonal forcing. Also we obatin the basic reproduction
	number
	\(R_0  = \frac{\beta _2 \mu _2 + \beta _1 \eta}{\mu _2(d + \alpha + \mu _1)}\)
	and
	\(R_v  = R_0\bigg[\frac{\mu _1 +\theta}{\mu _1 +\theta + \phi}\bigg]\).
	The system has endemic eqilibrium when \(R_v>1\) and \(R_v < 1\) with no
	positve endemic. We also obatin the critical values
	\(\phi = \frac{(\mu _1 +\theta)(\mu _2 \beta _2 +\beta _1 \eta)-\mu _2(\mu _1 + \theta)(d+\alpha +\mu _1)}{\mu _2(d+\alpha + \mu _1)} =\phi _v\)
	and
	\(\theta = \frac{\mu _1(\mu _2 \beta _2 +\beta _1 \eta)-\mu _2(\mu _1 + \phi)(d+\alpha +\mu _1)}{(\mu _2 \beta _2 +\beta _ \eta)-\mu _2(d+\alpha + \mu _1)} =\theta _v\).
	The model regards \(\phi\) and \(\theta\) as the control parameters,
	while the other parameters are fixed.
	
	\section{Results and Discussion}\label{primliminary-results}
	
	The study obtained the values for the parameters from literature to
	simulate the data. The parameter values are
	\(\mu _1 = 9.13*10^{-5},\mu_2 = 0.33,d = 0.013, a=0.2, b=0.15, \alpha=0.2, \alpha _1 =0.02, \alpha _2 =0.025, \eta =7, \theta =0.00002,\)
	and \(\phi =0.00001\). The unit for all these measurements is per day
	with the exception of contribution of infected individuals to the
	pathogen population. We now visualize the dynamics of the response and
	perform a sensitivity analysis of the critical values

	\begin{figure}[h!]
		
		{\centering \includegraphics[width=7in]{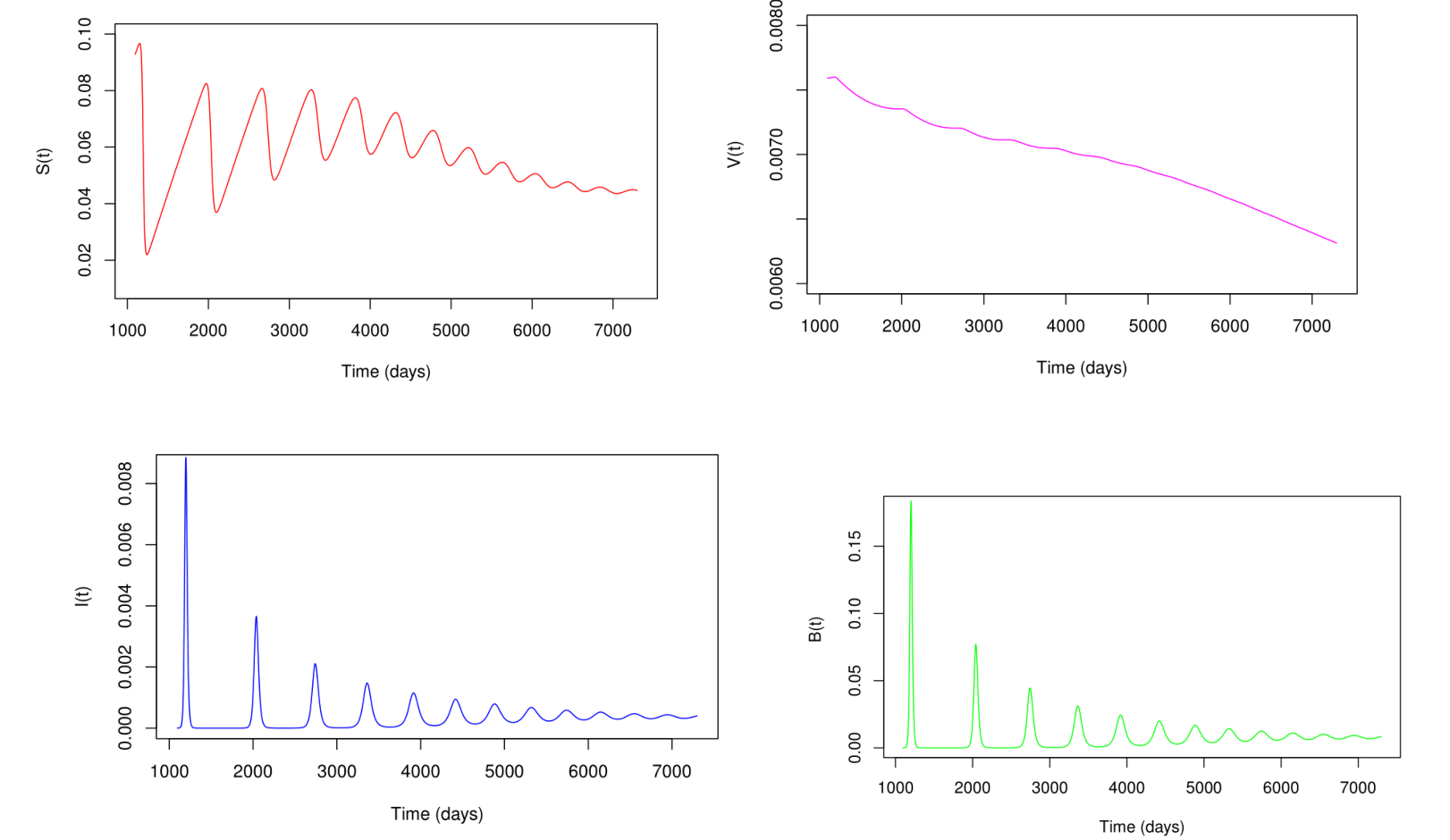} 
			
		}
		
		\caption{The curve of dynamics of the response variables forced by seasonal transmission and vaccination.}\label{fig:unnamed-chunk-1}
	\end{figure}

	\begin{figure}
		
		{\centering \includegraphics[width=7in]{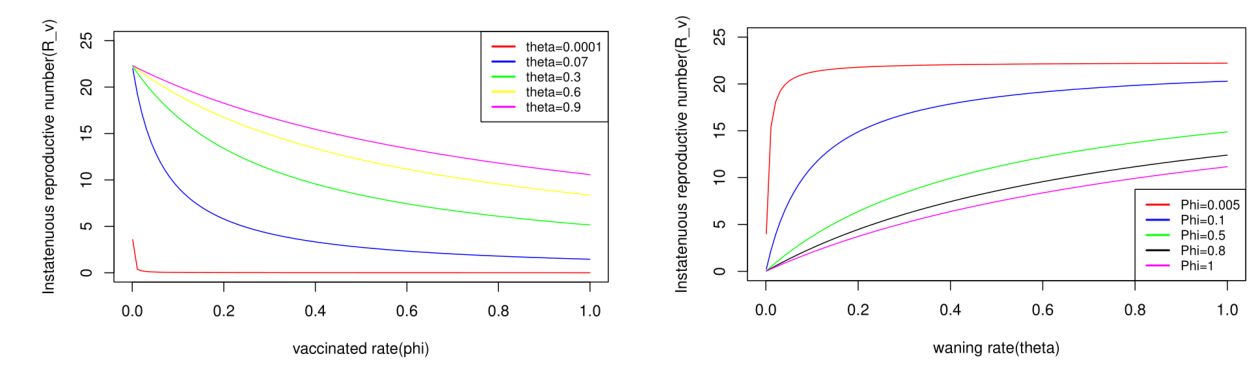} 
			
		}
		
		\caption{The curve of the control reproduction number with vaccinated rate(phi) when waning rate(theta) has some fixed value.}\label{fig:ses}
	\end{figure}

	From Figure \ref{fig:unnamed-chunk-1}, we observe a clear and pronounced relationship between the pathogen population size and the number of infected individuals over the simulated timeframe. Both curves exhibit remarkably similar temporal dynamics, underscoring the intrinsic connection between pathogen proliferation and the subsequent increase in host infection rates. Initially, there is a rapid rise in the pathogen population size, corresponding closely with an increase in the infected population. This trend aligns with established epidemiological expectations, as greater pathogen abundance generally increases exposure and transmission opportunities. The peak of the pathogen and infection curves, occurring nearly simultaneously, highlights the immediate and direct impact pathogen density exerts on cholera transmission dynamics.
	
	Subsequently, both curves show a pronounced decline shortly after reaching their respective peaks. This reduction can primarily be attributed to the short lifespan of the cholera pathogen, coupled with the effects of the implemented control measures—particularly vaccination. This rapid decline phase reflects an effective short-term control response, driven substantially by two critical parameters: the vaccination rate and the waning rate of vaccine-induced immunity. Detailed exploration of these parameters via sensitivity analysis (Figure \ref{fig:ses}) provides essential insights into how vaccine effectiveness can be strategically optimized.
	
	The sensitivity analysis distinctly reveals the critical influence these parameters exert on cholera transmission dynamics. As the vaccination rate increases, there is a marked reduction in the instantaneous reproduction number, clearly demonstrating the vaccine's pivotal role in controlling the spread of cholera. In contrast, increasing the waning rate of vaccine-induced immunity leads to an elevated , suggesting that faster loss of protective immunity could significantly undermine control efforts. These findings underscore the necessity of ensuring high vaccine efficacy and prolonged immunity duration to sustain effective cholera prevention. Therefore, vaccine formulations must be optimized for durability, and health authorities should consider implementing booster vaccination campaigns as integral components of public health strategies.
	
	The seasonal transmission dynamics depicted in the model provide further practical insights for public health interventions. These seasonal patterns suggest that strategic timing of vaccination programs, ideally preceding expected cholera peak seasons, could significantly enhance control outcomes. This recommendation is consistent with evidence from epidemiological studies advocating for targeted, seasonal vaccination campaigns to achieve optimal population-level protection.These detailed results corroborate and expand upon existing epidemiological literature {[}2{]}, reaffirming the critical role that vaccination strategies play in cholera outbreak control. Comprehensive cholera prevention requires the integration of vaccination with improved sanitation, hygiene education, and timely public health interventions. Through these combined efforts, communities can effectively manage and potentially eliminate cholera, substantially mitigating its public health impact.

	\section{Application}
	The outcomes of this study have direct applications in guiding public health decision-making, particularly in resource-limited settings where cholera outbreaks remain a recurrent challenge. By quantifying the effects of vaccination and waning immunity on the reproduction number, this model provides a robust analytical foundation for optimizing cholera vaccination programs. Health agencies and policy makers can apply these insights to allocate resources effectively, determine optimal vaccination schedules, and develop educational campaigns that stress the importance of timely immunization. Moreover, the seasonal nature of the disease dynamics identified in this model suggests that preemptive deployment of vaccines before high-transmission seasons could significantly mitigate outbreak severity. These applications are not limited to cholera alone; they also offer a framework adaptable to other waterborne and vaccine-preventable diseases.
	
	\section{Conclusion}
	This study provides a comprehensive mathematical framework for understanding cholera transmission dynamics under the influence of imperfect vaccination and seasonal forcing. The findings demonstrate that the vaccination rate and the rate of waning immunity critically determine the ability of public health systems to reduce the reproduction number below unity, thereby containing or eliminating the outbreak. The alignment of model predictions with known epidemiological behaviors enhances the credibility of these conclusions and affirms the potential of mathematical modeling as a decision-support tool. In light of these findings, it is imperative for public health programs to prioritize vaccine durability and optimize vaccination timing, particularly in regions experiencing seasonal cholera outbreaks. Future research should extend this model to include environmental interventions and stochastic variability to capture real-world complexities and improve predictive accuracy.

	\section*{References}\label{references}
	
	\begin{enumerate}
		\item McElroy, A. and Townsend, P. T. $(2009)$. Medical Anthropology in Ecological Perspective. Boulder, CO: Westview, $375$.
		
		\item Jingan, C., Zhanmin W., and Xueyong Z.Mathematical Analysis of a Cholera Model with Vaccination.Journal of Applied Mathematics Volume 2014, Article ID 324767, 16 pages.
		
		\item S. Liao and J. Wang, “Global stability analysis of epidemiological models based on Volterra-Lyapunov stable matrices,” Chaos, Solitons and Fractals, vol. 45, no. 7, pp. 966–977, 2012.
		
		\item C. T. Codec, “Endemic and epidemic dynamics of cholera: therole of the aquatic reservoir,” BMC Infectious Diseases, vol. 1, article 1, 2001.
		
		\item Y. Kang and H. Z. Zhang, “The study status of oral cholera vaccine,” Chinese Frontier Health Quarantine, vol. 28, no. s1, pp. 91–92, 2005.
		
		\item Lessler, J., Moore, S. M., Luquero, F. J., McKay, H. S., Grais, R., Henkens, M., ... and Azman, A. S. (2018). Mapping the burden of cholera in sub-Saharan Africa and implications for control: an analysis of data across geographical scales. The Lancet, 391(10133), 1908-1915.
		
		\item George, C. M., Monira, S., Sack, D. A., Rashid, M. U., Saif-Ur-Rahman, K. M., Mahmud, T., ... and Alam, M. (2016). Randomized controlled trial of hospital-based hygiene and water treatment intervention (CHoBI7) to reduce cholera. Emerging Infectious Diseases, 22(2), 233.
		
		\item Azman, A. S., Lessler, J., Luquero, F. J., Bhuiyan, T. R., Khan, A. I., and Sack, D. A. (2016). Estimating cholera incidence with cross-sectional serology. Science Translational Medicine, 8(361), 361ra154.
		
		\item Ali, M., Nelson, A. R., Lopez, A. L., and Sack, D. A. (2017). Updated global burden of cholera in endemic countries. PLoS Neglected Tropical Diseases, 11(6). 
		
		\item Lopez AL, Gonzales ML, Aldaba JG, Nair GB. Killed oral cholera vaccines: history, development and implementation challenges. Ther Adv Vaccines. 2014 Sep;2(5):123-36. doi: 10.1177/2051013614537819. PMID: 25177492; PMCID: PMC4144262.
		
		\item Bi Q, Ferreras E, Pezzoli L, Legros D, Ivers LC, Date K, Qadri F, Digilio L, Sack DA, Ali M, Lessler J, Luquero FJ, Azman AS; Oral Cholera Vaccine Working Group of The Global Task Force on Cholera Control. Protection against cholera from killed whole-cell oral cholera vaccines: a systematic review and meta-analysis. Lancet Infect Dis. 2017 Oct;17(10):1080-1088. doi: 10.1016/S1473-3099(17)30359-6. Epub 2017 Jul 17. PMID: 28729167; PMCID: PMC5639147.
		
	\end{enumerate}
	
\end{document}